\documentclass[%
 reprint,
 groupedaddress
 amsmath,amssymb,
 aps,
]{revtex4-2}
\usepackage{graphicx}
\usepackage{dcolumn}
\usepackage{bm}
\usepackage{hyperref}
\usepackage{booktabs}
\usepackage{multirow}
\usepackage{amsmath}
\usepackage{colortbl}
\usepackage{color}
\hypersetup{hypertex=true,
            colorlinks=true,
            linkcolor=blue,
            anchorcolor=blue,
            citecolor=blue}

\begin{document}

\preprint{APS/123-QED}

\title{Nuclear level densities in the relativistic Hartree-Bogoliubov plus combinatorial framework}

\author{Pengxiang Du}

\author{Jian Li}
\email{jianli@jlu.edu.cn}
\affiliation{College of Physics, Jilin University, Changchun 130012, China}

\date{\today}

\begin{abstract}
A systematic study of nuclear level densities has been carried out within the relativistic Hartree-Bogoliubov plus combinatorial framework. Calculations were performed for even-even nuclei with available experimental data, based on the relativistic energy density functionals DD-ME2, DD-PC1, and PC-PK1. The overall performance of the model is assessed against experimental data. On this basis, the effects of different functionals, pairing correlations, deformation, and other relevant factors on nuclear level densities are examined. The results show that the present framework provides a good description of the experimental level density, and reproduces the 
$s$-wave neutron resonance spacings with an accuracy comparable to that of the best existing global models. Furthermore, differences among the adopted relativistic density functionals in the nucleon effective mass at saturated nuclear matter are transmitted to the predicted level densities and constitute the main source of the differences among the results obtained with the three functionals.

\end{abstract}

\maketitle

\section{\label{sec:1}introduction}

From astrophysical nucleosynthesis to nuclear data evaluation and nuclear energy applications, reliable reaction calculations are closely connected with the nuclear level density (NLD). In statistical approaches such as the Hauser-Feshbach formalism, NLD is an essential input that directly influences calculated compound-nucleus cross sections, including capture and fission~\cite{hauser1952inelastic,rajasekaran1981nuclear,Rauscher1997nuclear,goriely2008improved}. Despite its importance, experimental information on NLD remains limited to a restricted range of nuclei and excitation energies.

The theoretical study of NLD can be traced back to Bethe’s pioneering work based on the Fermi-gas picture~\cite{bethe1936attempt,bethe1937nuclear}. Subsequently, as shell, pairing, and collective effects were introduced to account for experimental observations, a variety of models were proposed, such as the back-shifted Fermi-gas model (BFM)~\cite{dilg1973level} and the generalized superfluid model (GSM)~\cite{ignatyuk1975phenomenological}, among others. Meanwhile, the constant-temperature model (CTM)~\cite{gilbert1965composite} can well describe the approximately exponential behavior of NLDs over a certain excitation-energy range, and this feature has also been reflected in Oslo data in recent years~\cite{Guttormsen2015Experimental,Goriely2022Comprehensive}, so that it has continued to attract attention in NLD studies~\cite{zelevinsky2019nuclear}. These models can reproduce experimental data simply and effectively through parameter adjustment~\cite{plyaskin2000level,von2005systematics,koning2008global,capote2009ripl}, and have therefore been widely used. However, when applied to nuclei far from stability or to regions where experimental information is scarce, such as those relevant to nuclear astrophysics, the above approaches often become less reliable. Therefore, compared with phenomenological models that rely on empirical parameter adjustments, approaches based on a microscopic description of nuclear structure are generally to be preferred.

Current microscopic studies of NLD mainly follow two routes: the shell model and mean-field approaches. In general, the conventional shell model is limited in NLD studies by the huge configuration space and the computational complexity associated with its diagonalization. To reduce the computational cost while keeping the main physical features of the shell model, the moments method~\cite{senkov2010high,senkov2013high,senkov2016nuclear}, the stochastic estimation method~\cite{shimizu2016stochastic,chen2023shell}, the extrapolated Lanczos matrix method~\cite{ormand2020microscopic}, and related approaches have been developed and mainly applied to light and medium-mass nuclei. In addition, the projected shell model based on multi-quasiparticle configurations has also been used for NLD calculations~\cite{wang2023projected}. The shell-model Monte Carlo method provides a statistical description of NLD through thermodynamic quantities and the partition function, and has been applied to studies of lanthanide nuclei~\cite{ormand1997estimating,alhassid1999particle,white2000shell,alhassid2003nuclear,Alhassid2007spim,Alhassid2015Direct,Alhassid2021Nuclear}, but its applicability is constrained by the sign problem. Similar ideas have also been widely used in mean-field approaches, where the intrinsic level density is usually extracted through the inverse Laplace transform of the partition function, and collective effects are then introduced on this basis. Compared with the shell model, the statistical method based on mean-field models can be more readily extended to heavy nuclei and has therefore been widely used in NLD studies across the nuclear chart~\cite{choudhury1977nuclear,goriely1996new,demetriou2001microscopic,Bezbakh2014level,zhao2020microscopic,zhang2023level}. 

Another widely used approach within the mean-field framework is the combinatorial method~\cite{hilaire2001combinatorial}, which employs a state generating function constructed from single-particle levels to perform a direct counting of particle-hole (p-h) excitations. In contrast to statistical method, this method yields spin- and parity-dependent NLDs and captures the non-statistical behavior at low excitation energies. Such features can have a significant impact on cross section predictions~\cite{Goko2006partial}. Under the assumption of axial symmetry, global NLD studies based on the Skyrme Hartree-Fock-Bogoliubov plus combinatorial method~\cite{hilaire2006global,goriely2008improved1} as well as the temperature-dependent Gogny Hartree-Fock-Bogoliubov plus combinatorial method~\cite{hilaire2012temperature} have been reported. More recently, the triaxial Hartree-Fock-Bogoliubov plus combinatorial method has also been proposed for the description of NLDs~\cite{Goriely2026Improved}. These results agree with experimental data at a level comparable to phenomenological models, showing the considerable practical value of the combinatorial method.

Over the past decades, relativistic energy density functional (EDF) theory has become an important microscopic theoretical tool for describing both ground-state and excited properties of nuclei~\cite{RING1996Relativistic,NIKSIC2011Relativistic,meng2016relativistic,Li2018Nuclear,Xie2024impact,Du2024exploring,SHANG2024pseudospin,zang2025coexistence}. Within this framework, the combinatorial method was first applied to NLD calculations in Ref.~\cite{geng2023calculation}. Subsequently, in Ref.~\cite{jiang2024nuclear}, pairing correlations were included through the relativistic Hartree-Bogoliubov (RHB) theory, and the Strutinsky method was used to smooth out fluctuations in the NLD at low excitation energies, resulting in a successful description of the NLD for $^{\mathrm{112}}\mathrm{Cd}$. More recently, the NLDs of Sn isotopes were also investigated~\cite{He2026Systematic}. These studies demonstrate the feasibility of calculating NLDs within the combinatorial method based on relativistic EDF theory. However, relevant investigations remain limited, and systematic studies over a broader range of nuclei are still needed to assess the overall performance of this method. Furthermore, in such calculations, the influence of different relativistic density functionals, as well as key physical factors such as pairing correlations and deformation, on NLDs still needs to be clarified. Therefore, a systematic study within the RHB plus combinatorial framework is necessary. 

In this work, the RHB plus combinatorial method based on several relativistic density functionals is applied to NLD calculations for nearly 70 even-even nuclei with available experimental data. The main features of the results and the influence of relevant physical factors are analyzed, and the overall performance of the calculations is assessed by comparison with experimental data.
The article is arranged as follows. In Sec.~\ref{sec:2}, the theoretical frameworks of the RHB theory and the combinatorial method, together with the numerical details adopted in the calculations, are introduced. The calculated results and corresponding discussion are presented in Sec.~\ref{sec:3}. Finally, Sec.~\ref{sec:4} contains the summary.

\section{\label{sec:2}Relativistic Hartree-Bogoliubov theory and the combinatorial method
}

\subsection{The relativistic Hartree-Bogoliubov theory}

The relativistic Hartree-Bogoliubov theory provides a self-consistent description of the nuclear ground-state, with the mean field and pairing correlations treated in a unified way~\cite{VRETENAR2005Relativistic,MENG2006Relativisti,NIKSIC2014DIRHB}. Within this framework, the nuclear system is described by the RHB equation, which reads
\begin{equation}
\begin{pmatrix}
\hat{h}_{D}-\lambda_{\tau} & \hat{\Delta} \\
-\hat{\Delta}^{*} & -\hat{h}_{D}^{*}+\lambda_{\tau}
\end{pmatrix}
\begin{pmatrix}
U_{k} \\
V_{k}
\end{pmatrix}
=
E_{k}
\begin{pmatrix}
U_{k} \\
V_{k}
\end{pmatrix},
\label{eq:rhb}
\end{equation}
where $\hat{h}_{D}$ is the single-nucleon Dirac Hamiltonian, $\lambda_{\tau}$ ($\tau=n,p$) are the
Fermi energy for neutrons and protons, respectively, $\hat{\Delta}$ represents the
pairing potential, $U_{k}$ and $V_{k}$ are the quasiparticle
wave functions, and $E_{k}$ is the quasiparticle energy. The single-nucleon Dirac Hamiltonian $\hat{h}_{D}$ is given by
\begin{equation}
\hat{h}_{D}
=
\boldsymbol{\alpha}\cdot\boldsymbol{p}
+\beta\left[m+S\right]
+V,
\end{equation} 
where $S$ and $V$ denote the attractive scalar and repulsive vector potentials, respectively. In the present work, three different types of relativistic EDFs are adopted: the density-dependent meson-exchange functional DD-ME2~\cite{Lalazissis2005new}, the density-dependent point-coupling functional DD-PC1~\cite{Nikifmmode2008Relativistic}, and the nonlinear point-coupling functional PC-PK1~\cite{Zhao2010new}. Accordingly, the explicit forms of the corresponding potentials are different. For convenience of discussion, the PC-PK1 functional is taken as an example below to present the explicit expressions:
\begin{align}
S &=
\alpha_{S}\rho_{S}
+\beta_{S}\rho_{S}^{2}
+\gamma_{S}\rho_{S}^{3}
+\delta_{S}\Delta\rho_{S},
\\
V &= \alpha_{V}\rho_{V}
   +\gamma_{V}\rho_{V}^{3}
   +\delta_{V}\Delta\rho_{V} \notag\\
  &{}+ \tau_{3}\alpha_{TV}\rho_{TV}
   +\tau_{3}\delta_{TV}\Delta\rho_{TV}
   +eA_{0}\frac{1-\tau_{3}}{2}.
\end{align}
Here, $\rho_{S}$, $\rho_{V}$, and $\rho_{TV}$ are the scalar, vector, and isovector-vector densities, respectively. The derivative terms simulate finite-range effects, and the last term in $V$ represents the Coulomb interaction for protons. The pairing potential $\hat{\Delta}$ is written as
\begin{equation}
\hat{\Delta}_{n_{1}n_{2}}
=
\frac{1}{2}
\sum_{n_{1}^{\prime}n_{2}^{\prime}}
\langle n_{1} n_{2} | V^{pp} | n_{1}^{\prime} n_{2}^{\prime} \rangle
\kappa_{n_{1}^{\prime}n_{2}^{\prime}},
\end{equation}
where $V^{pp}$ denotes the pairing interaction, and $\kappa$ is the pairing tensor. In the present work, $V^{pp}$ is taken in a separable form. This interaction is constructed to reproduce the pairing properties of the Gogny force in the $^{1}S_{0}$ channel in nuclear matter, and its explicit form can be found in Ref.~\cite{TIAN2009Afinite,tian2009Separable}.

The RHB equation~\eqref{eq:rhb} is solved in 16 harmonic-oscillator shells, and the convergence has been checked. A pairing strength of $G=728~\mathrm{MeV\,fm^{3}}$ is adopted. The resulting self-consistent single-particle structure and pairing information are then used in the subsequent combinatorial calculation.

\subsection{The combinatorial method}

In the combinatorial method, the NLD is obtained by counting incoherent p-h excitations and taking collective effects into account~\cite{hilaire2001combinatorial,hilaire2006global,goriely2008improved1,hilaire2012temperature,Goriely2026Improved}. The incoherent p-h excitations are constructed from the single-particle levels given by the RHB calculation. Their combinatorial information is encoded in the generating function given below:
\begin{equation}
\mathcal{Z}(x_{1},x_{2},x_{3},x_{4},y,t)
=
\prod_{k=1}^{4}\prod_{i=1}^{I_{k}}
\left(1+x_{k}p_{i}^{k}y^{\epsilon_{i}^{k}}t^{m_{i}^{k}}\right),
\end{equation}
where $k$ labels the four single-particle subspaces (neutron particles, neutron holes, proton particles, and proton holes), and $I_k$ is the total number of single-particle states in subspace $k$. The variable $x_k$ tracks the number of single-particle states involved in the combination in subspace $k$, while $y$ and $t$ track the excitation energy and spin projection, respectively. The quantities $p_i^k$, $\epsilon_i^k$, and $m_i^k$ denote the parity, energy, and spin projection of single-particle state $i$ in subspace $k$. After expanding the product, every generated term corresponds to p-h excitation states with definite excitation energy, spin projection, and parity. The coefficient of each term directly gives the exact combinatorial number of such states. This counting is performed on a grid with equal energy spacing ($\epsilon_{0}$). In evaluating the excitation energies of these p-h configurations, the ground-state pairing gap is applied to generated hole states. While this treatment neglects the quenching of the pairing field at higher excitation energies, it serves as an effective phenomenological compensation for the missing residual interactions, preventing an overestimation of the state densities at low and medium energies.

Based on incoherent p-h excitations, vibrational effects are taken into account by folding phonon states. The phonon states are treated statistically using a generalized boson partition function, and the quadrupole, octupole, and hexadecapole vibrational modes are included. The choice of phonon energies and the detailed procedure follow those of Ref.~\cite{goriely2008improved1}. In the present work, the maximum number of folded phonons is taken to be three for light and medium-mass nuclei. For rare-earth nuclei around $A\sim150$ and heavier systems, the maximum number is increased to four in order to account more fully for their potentially richer collective vibrational couplings. Although the existence of pure three-phonon excitations in low-lying collective states remains under debate~\cite{Aprahamian1987first,Batchelder2012low}, the phonon-number cutoff adopted in this work remains an empirical modeling choice. Given the damping of collective effects at high excitation energies, the exciton number of the p-h states folded with phonons is restricted to at most four~\cite{goriely2008improved1,BANERJEE2017Direct,SANTHOSH2023Experimental}. 

With the phonon states folded in, the state density can then be obtained as
\begin{equation}
\omega(U,M,P)=\frac{C(U,M,P)}{\epsilon_0},
\end{equation}
where $C(U,M,P)$ denotes the number of states with a given excitation energy $U$, spin projection $M$, and parity $P$, and the $\epsilon_{0}$ is the energy step size introduced above. The state density is inevitably affected by the adopted energy step size. An excessively large $\epsilon_{0}$ may lead to an overestimation of the number of p-h combinations, and this effect is particularly pronounced for heavy nuclei. In the present work, $\epsilon_{0}=0.01$ MeV is adopted. In addition, the state density exhibits noticeable fluctuations in the low excitation energy region. To remove the fluctuations, the smoothing strategy of Ref.~\cite{jiang2024nuclear} is followed, while a simpler Gaussian kernel is used here. The smoothed state density is written as
\begin{equation}
\label{eq:formula_smooth}
\tilde{\omega}(U,M,P)
=
\int \omega(U',M,P)\,
\frac{1}{\sigma\sqrt{2\pi}}
e^{-\frac{(U-U')^{2}}{2\sigma^{2}}}
\, dU',
\end{equation}
where $\sigma$ is the smoothing width and is taken as $0.2$ MeV.

When the nucleus under consideration displays spherical symmetry, the NLD for a given excitation energy $U$, spin $J$, and parity $P$ can be obtained through the relation
\begin{equation}
\rho(U,J,P)
=
\tilde\omega(U,M=J,P)-\tilde\omega(U,M=J+1,P).
\end{equation}
For axially symmetric deformed systems, the rotational effect must be treated explicitly, and the final NLD can be expressed as
\begin{equation}
\begin{alignedat}{2}
\rho(U,J,P) &= \frac{1}{2}
\sum_{M=-J,\;M\neq 0}^{J}
\tilde\omega\!\left(U-E_{\mathrm{rot}}^{J,M},M,P\right) \\
            &+ \delta_{(J=\mathrm{even})}\delta_{(P=+)}
\tilde\omega\!\left(U-E_{\mathrm{rot}}^{J,0},0,P\right) \\
            &+ \delta_{(J=\mathrm{odd})}\delta_{(P=-)}
\tilde\omega\!\left(U-E_{\mathrm{rot}}^{J,0},0,P\right).
\end{alignedat}
\end{equation}
The rotation energy $E^{J,M}_\mathrm{rot}$ is obtained from
\begin{equation}
E_{\mathrm{rot}}^{J,M}
=
\frac{\hbar^{2}}{2\mathcal{J}_{\perp}}
\left[ J(J+1)-M^{2} \right],
\end{equation}
where $\mathcal{J}_{\perp}$ denotes the moment of inertia for rotation about an axis perpendicular to the symmetry axis, and it is calculated using the Inglis-Belyaev formula~\cite{Inglis1956Nuclear,BELIAEV1961Concerning}. In the present work, the NLDs are calculated up to $J=50\hbar$. Additionally, the smoothed state density $\tilde{\omega}$ will be denoted simply by $\omega$ in the following discussion.

\section{\label{sec:3}Results and discussion}

\subsection{Calculated NLDs versus Oslo data and origin of the differences among EDFs}

\begin{figure*}
\centering
\includegraphics[width=18cm]{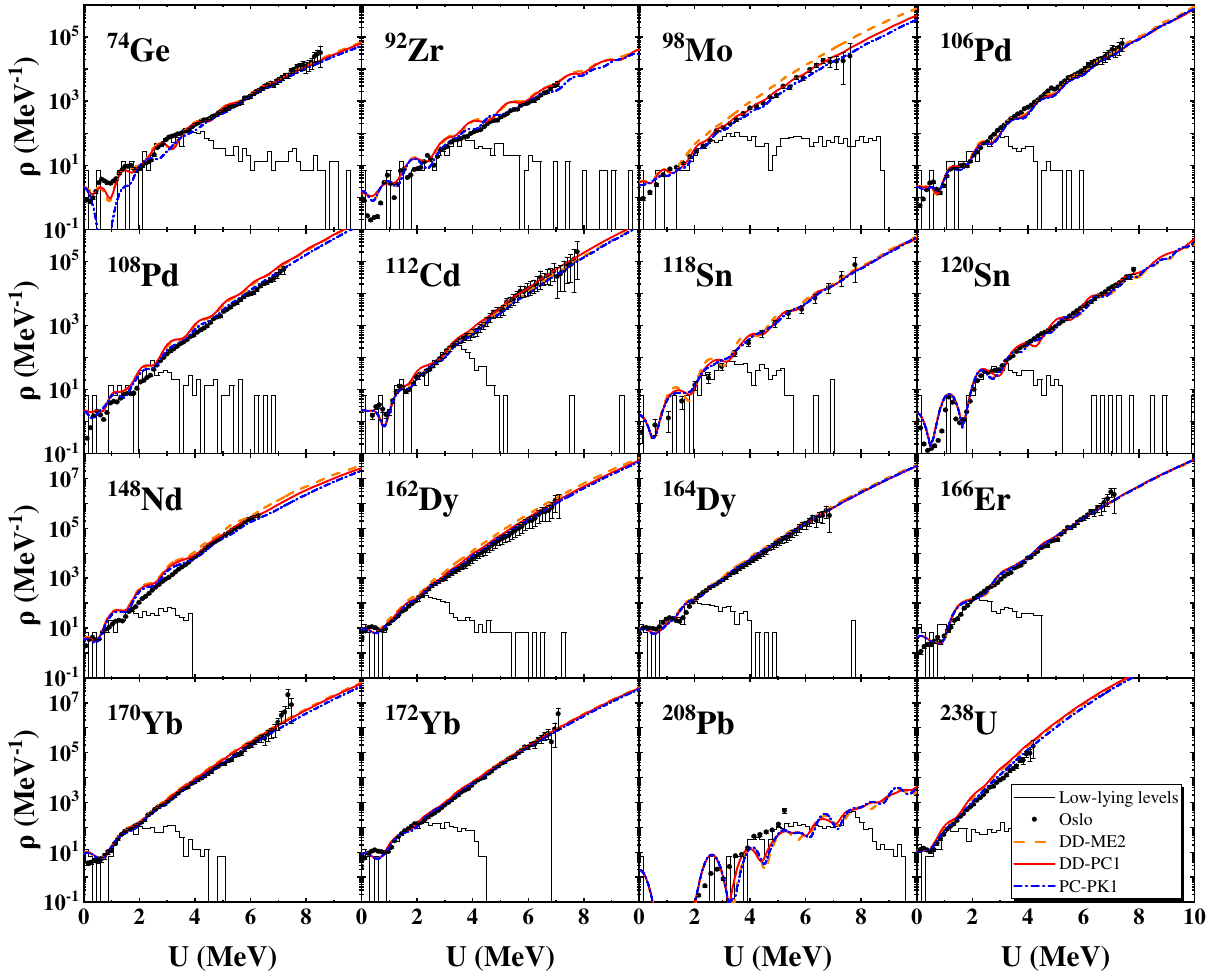}
\caption{\label{fig:ld}(Color online) Comparison of NLDs calculated with the RHB plus combinatorial method and experimental data. The black solid line denotes the NLD derived from known low-lying levels, and the black dots represent the Oslo data~\cite{Renstrom2016Low,Guttormsen2017Quasicontinuum,Utsunomiya2013Photoneutron,Eriksen2014Pygmy,Goriely2022Comprehensive,Toft20101Level,Markova2022Nuclear,GUTTORMSEN2021Strong,Renstrom2018Verification,Melby2001Thermal,Agvaanluvsan2004Level,Syed2009Level,Guttormsen2013Constant}. The yellow dashed line shows the result based on DD-ME2, while the red solid line and the blue dash-dotted line represent the results obtained with DD-PC1 and PC-PK1, respectively.}
\end{figure*}

As mentioned above, experimental data on NLDs remain rather scarce, and the Oslo method is one of the important sources of experimental NLD information in the low and intermediate excitation energy region~\cite{schiller2000extraction,MIDTBO2021a}. Figure~\ref{fig:ld} shows a comparison between the NLDs calculated with the three adopted relativistic EDFs and the experimental data from the Oslo group for 16 nuclei ranging from $^{74}\mathrm{Ge}$ to $^{238}\mathrm{U}$~\cite{Renstrom2016Low,Guttormsen2017Quasicontinuum,Utsunomiya2013Photoneutron,Eriksen2014Pygmy,Goriely2022Comprehensive,Toft20101Level,Markova2022Nuclear,GUTTORMSEN2021Strong,Renstrom2018Verification,Melby2001Thermal,Agvaanluvsan2004Level,Syed2009Level,Guttormsen2013Constant}. In addition, the available experimental low-lying levels are displayed in the form of NLDs. It should be noted that the NLDs extracted by the Oslo method still contain some degree of model dependence, mainly associated with the extraction and normalization~\cite{Larsen2011Analysis,MIDTBO2021a}. Accordingly, model-dependent renormalization of the Oslo data has been adopted in the literature when different NLD models are compared on the same footing~\cite{Goriely2022Comprehensive}. In the present work, however, such a renormalization is not adopted, since the aim is to retain a common experimental reference and to illustrate, within the same framework, the differences arising solely from the use of different functionals. For $^{112}\mathrm{Cd}$ and $^{120}\mathrm{Sn}$, the Oslo data are additionally constrained by the shape method, which extracts model-independent NLD data by exploiting the internal physical consistency of the primary experimental data~\cite{wiedeking2021independent,mucher2023Extracting}. As can be seen from Fig.~\ref{fig:ld}, the three adopted relativistic EDFs reproduce the experimental NLD data well. At low excitation energies, the calculated results are broadly consistent with the known low-lying levels. Within the range covered by the Oslo data, the calculations capture both the approximately exponential increase of the experimental NLDs and, to some extent, the observed nonstatistical features. In particular, for the shape method results of $^{112}\mathrm{Cd}$ and $^{120}\mathrm{Sn}$, the present calculations provide a good description. 

\begin{figure*}
\centering
\includegraphics[width=18cm]{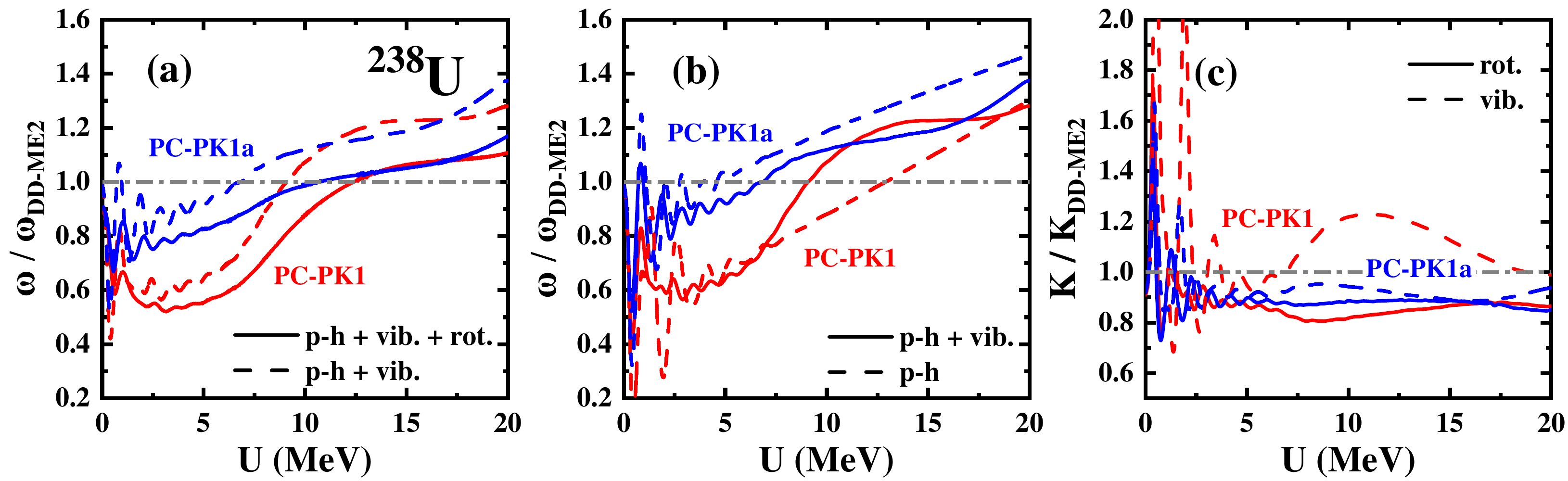}
\caption{\label{fig:mass1}Comparison of the results from PC-PK1 (red) and the adjusted PC-PK1, denoted by PC-PK1a (blue), with those from DD-ME2 (black) for $^{\mathrm{238}}\mathrm{U}$ at the constrained deformation ${\beta}_2=0.3$. (a) Ratios of the state densities calculated with PC-PK1 and PC-PK1a to those calculated with DD-ME2 as functions of excitation energy. The solid lines correspond to the state densities including p-h, vibrational, and rotational contributions, while the dashed lines correspond to those including p-h and vibrational contributions only. 
(b) Same as panel (a), but for the comparison between the state densities including p-h and vibrational contributions (solid lines) and the p-h state densities only (dashed lines).  (c)  Ratios of the rotational enhancement $\mathrm{K}_{\mathrm{rot}}$ (solid lines) and the vibrational enhancement $\mathrm{K}_{\mathrm{vib}}$ (dashed lines) obtained with PC-PK1 and PC-PK1a to the corresponding values obtained with DD-ME2.}\label{fig:mass1}
\end{figure*}

In general, the NLDs calculated with the three relativistic EDFs show a high degree of consistency, but two types of differences can still be identified, namely differences in curve shape and in absolute magnitude. The former mainly appear in a few nuclei and are primarily related to the different ground-state deformations predicted by the functionals. Taking $^{74}\mathrm{Ge}$ as an example, PC-PK1 predicts a prolate ground-state with a quadrupole deformation parameter $\beta_2=0.10$, whereas the other two EDFs give an oblate shape with $\beta_2=-0.20$. The latter, by contrast, exhibit a more general systematic behavior. Within the excitation-energy range shown in Fig.~\ref{fig:ld}, PC-PK1 usually yields the smallest NLDs, and in most cases the ordering is DD-ME2 $>$ DD-PC1 $>$ PC-PK1. This rather general ordering reflects systematic differences in the underlying single-particle structure, which are closely related to the effective mass. A larger effective mass tends to correspond to a higher single-particle level density around the Fermi surface and, in a Fermi-gas picture, to a larger level-density parameter $a$, thus favoring a larger overall magnitude of the NLD~\cite{PRAKASH1983Effective,SHLOMO1992Energy}. Contrary to this overall trend, however, PC-PK1 has the largest Dirac mass $M_D^{*}$ and Landau mass $M_L^{*}$ at nuclear matter saturation, whereas DD-ME2 has the smallest. The Dirac mass mainly reflects the scalar potential, while the Landau mass is more directly related to the single-particle level density around the Fermi surface.

\begin{table*}
\caption{\label{tab:edf_comparison}%
Comparison of the Dirac and Landau effective masses, $M_D^{*}/M$ and $M_L^{*}/M$, at nuclear matter saturation for DD-ME2, PC-PK1, and several adjusted versions of PC-PK1. For the same set of functionals, the neutron and proton single-particle level densities near the Fermi surface, $g_{\varepsilon_f}$ (in $\mathrm{MeV}^{-1}$), at different smoothing scales, as well as the average pairing gaps $\langle uv\Delta\rangle$ (in $\mathrm{MeV}$), are obtained from constrained RHB calculations for $^{238}\mathrm{U}$ at $\beta_{2}=0.3$.
}
\begin{ruledtabular}
\begin{tabular}{ccccccccccc}
\multirow{2}{*}{\textrm{EDF}}
& \multirow{2}{*}{$M_D^{*}/M$}
& \multirow{2}{*}{$M_L^{*}/M$}
& \multicolumn{2}{c}{$g_{\varepsilon_f}^{\sigma=0.5}$}
& \multicolumn{2}{c}{$g_{\varepsilon_f}^{\sigma=1.0}$}
& \multicolumn{2}{c}{$g_{\varepsilon_f}^{\sigma=2.0}$}
& \multicolumn{2}{c}{$\langle uv\Delta\rangle$} \\
\cline{4-5}
\cline{6-7}
\cline{8-9}
\cline{10-11}
& &
& $n$ & $p$
& $n$ & $p$
& $n$ & $p$
& $n$ & $p$ \\
\colrule
DD-ME2          & 0.572 & 0.635 & 4.043 & 4.046 & 3.541 & 4.067 & 4.685 & 3.985 & 0.513 & 0.923 \\
PC-PK1          & 0.591 & 0.652 & 4.493 & 3.497 & 3.998 & 4.061 & 5.193 & 4.170 & 0.704 & 1.053 \\
PC-PK1a       & 0.586 & 0.648 & 4.425 & 4.153 & 3.860 & 4.140 & 5.001 & 4.094 & 0.658 & 1.032 \\
PC-PK1b       & 0.582 & 0.644 & 4.288 & 4.469 & 3.774 & 4.137 & 4.885 & 4.029 & 0.626 & 1.015 \\
PC-PK1c       & 0.579 & 0.643 & 4.193 & 4.640 & 3.733 & 4.137 & 4.828 & 3.997 & 0.611 & 1.007 \\
\end{tabular}
\end{ruledtabular}
\end{table*}

To examine the above situation in more detail, Fig.~\ref{fig:mass1} shows the ratios of the state densities and related quantities calculated with PC-PK1 and DD-ME2 for $^{238}\mathrm{U}$ at the constrained deformation $\beta_2 = 0.30$. It should be noted that, when both vibrational and rotational effects are taken into account, the total state density is obtained from the NLD by including the angular-momentum degeneracy, namely $\sum_{J}(2J+1)\rho(U,J)$. The results obtained with an adjusted version of PC-PK1, denoted by PC-PK1a, are also shown. Compared with the original PC-PK1, the two parameters $\alpha_S$ and $\alpha_V$ in PC-PK1a are scaled by factors of 1.015 and 1.016, respectively. Here, $\alpha_S$ and $\alpha_V$ denote the coupling strengths in the attractive isoscalar-scalar and repulsive isoscalar-vector channels of the point-coupling interaction. By this readjustment, the effective masses in PC-PK1a are reduced and become closer to those of DD-ME2. In Fig.~\ref{fig:mass1}(a), the state density from PC-PK1 is significantly smaller than that from DD-ME2 at low excitation energies and exhibits pronounced oscillations, whereas the PC-PK1a result stays systematically closer to DD-ME2 over the whole energy range. After removing the rotational contribution, the ratios for both functionals shift upward, with a more pronounced change for PC-PK1 at higher energies, especially around 12--16~MeV. This indicates a weaker rotational contribution in PC-PK1. Figure~\ref{fig:mass1}(b) further reduces the comparison to the cases with and without vibrational contributions. For the pure p-h excitations, the ratio obtained with PC-PK1 is already below unity at low excitation energies, but rises more rapidly with increasing energy than that obtained with PC-PK1a, indicating that the difference already exists in the underlying p-h background. The vibrational contribution reshapes the energy dependence such that the ratio rises more rapidly at first and then tends to level off, with this additional effect gradually fading away toward 20~MeV. In contrast, for PC-PK1a, including the vibrational contribution shifts the ratio downward, indicating a weaker vibrational enhancement than in DD-ME2. In addition, both the rotational and vibrational contributions smooth the oscillatory behavior in the low-energy region, where the oscillations themselves reflect differences in the non-statistical features. Figure~\ref{fig:mass1}(c) shows the corresponding rotational and vibrational enhancements, obtained from the ratios between the state densities calculated with and without rotational or vibrational contributions. Below 5~MeV, strong oscillations are seen for PC-PK1, whereas those of PC-PK1a are much weaker and remain closer to the DD-ME2 result over the whole energy range.

The trends shown in Fig.~\ref{fig:mass1} can be further understood from Table~\ref{tab:edf_comparison}, which lists the effective masses, the single-particle level densities around the Fermi surface, and the average pairing gaps for DD-ME2, PC-PK1, and PC-PK1a. The single-particle level density around the Fermi surface is defined as
\begin{equation}
g_{\varepsilon_f}^{\sigma}
=
\frac{1}{\sqrt{2\pi}\sigma}
\int_{-\infty}^{\infty}
\sum_i \delta(\varepsilon-\varepsilon_i)\,
e^{-\frac{(\varepsilon-\varepsilon_f)^2}{2\sigma^2}}
\, d\varepsilon,
\end{equation}
where $\varepsilon_f$ is the Fermi energy, $\sigma$ is the Gaussian smoothing width, and $\varepsilon_i$ denotes the discrete single-particle energies. The average pairing gap is defined as
\begin{equation}
\langle uv\Delta\rangle
=
\frac{\sum_k u_k v_k \Delta_k}{\sum_k u_k v_k},
\end{equation}
where $u_k$ and $v_k$ are the quasiparticle amplitudes, and $\Delta_k$ is the pairing gap associated with the single-particle state $k$. Over a broad energy range around the Fermi surface ($\sigma = 2.0$~MeV), PC-PK1 gives a denser single-particle spectrum and a larger average pairing gap. Consequently, the larger $g_{\varepsilon_f}$ at this scale tends to accelerate the growth of p-h states at higher excitation energies. On a more local scale ($\sigma = 0.5, 1.0$~MeV), however, the smaller proton $g_{\varepsilon_f}$ suppresses the low-energy p-h excitations, and the larger pairing gap further strengthens this suppression. PC-PK1a has smaller $g_{\varepsilon_f}$ and average pairing gaps, so that its p-h state density is closer to DD-ME2 at low excitation energies and grows more slowly at higher energies. As for the vibrational contribution, it is not described self-consistently from a unified microscopic theory in the present framework, but is largely governed by the size of the p-h space and its growth with excitation energy. When the p-h space is sparse but grows rapidly, the vibrational correction can account for a larger fraction of the state density, as in PC-PK1. When the p-h space is denser, its relative effect is diluted more quickly, as in PC-PK1a. The rotational contribution depends on both the state density after vibrational folding and the moment of inertia. For PC-PK1, the larger effective mass is accompanied by stronger pairing, which suppresses low-energy excitations and reduces the moment of inertia, thereby leading to a weaker rotational contribution.

\begin{figure}
\centering
\includegraphics[width=8cm]{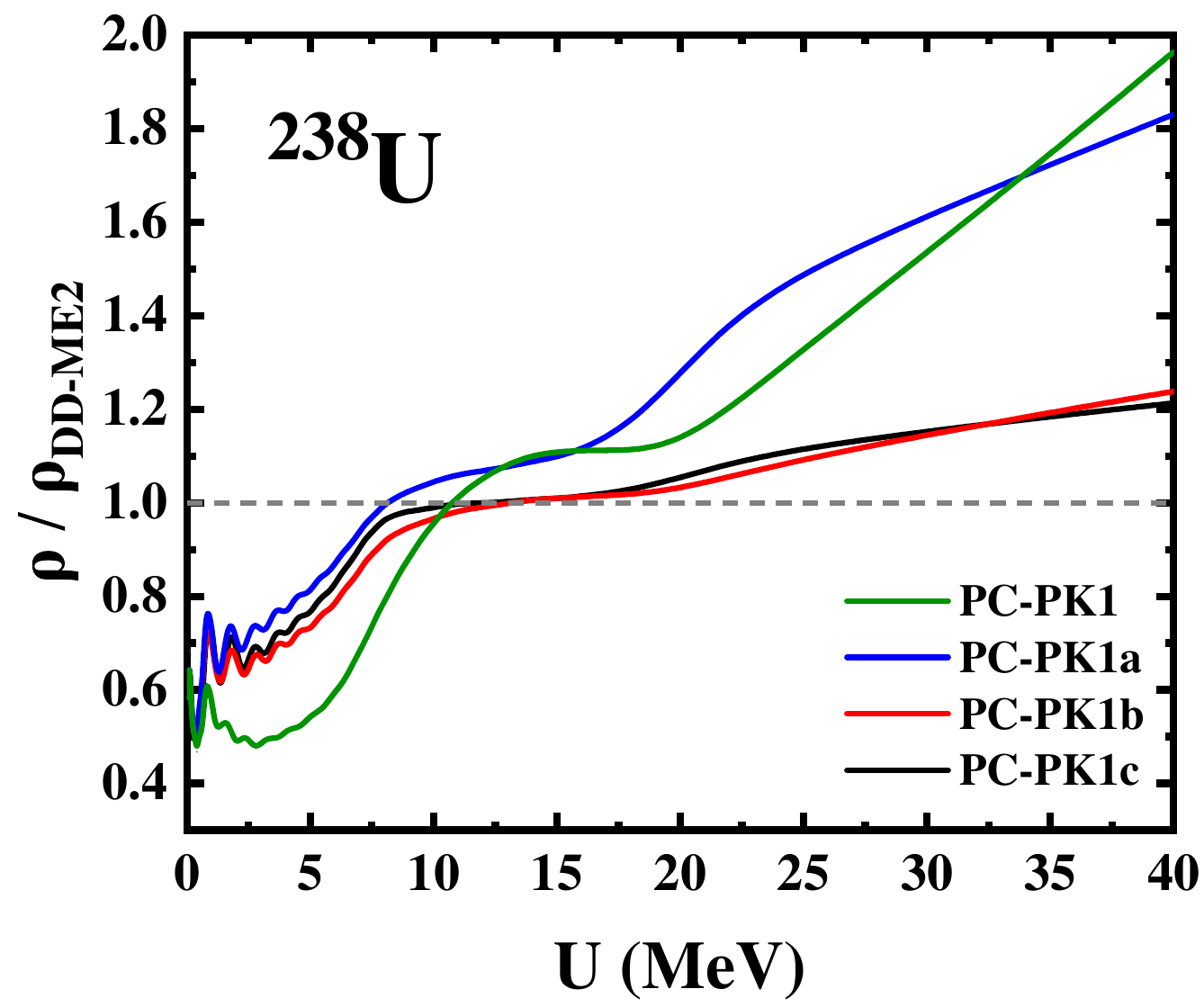}
\caption{\label{fig:mass2}(Color online) Ratios of the NLDs of $^{\mathrm{238}}\mathrm{U}$ at ${\beta}_2=0.3$ obtained with PC-PK1 and its adjusted versions to those obtained with DD-ME2, with the Dirac mass $M_D^{*}/M$ varying from $0.591$ to $0.579$}\label{fig:mass2}
\end{figure}

The above results indicate that the state-density differences between PC-PK1 and DD-ME2 arise from the competition among the single-particle level density around the Fermi surface, pairing correlations, and collective contributions. As $\alpha_S$ and $\alpha_V$ are readjusted, these relevant quantities move systematically toward the DD-ME2 values, so that the state densities obtained with PC-PK1a also become closer to those of DD-ME2. To examine this trend, $\alpha_S$ is fixed at the PC-PK1a value, while $\alpha_V$ is further adjusted to 1.014 and 1.013 times its original PC-PK1 value, leading to PC-PK1b and PC-PK1c. This readjustment enhances the relative role of the scalar self-energy, binds the nucleons more deeply, and consequently brings the Dirac and Landau mass closer to that of DD-ME2. The corresponding effective masses, $g_{\varepsilon_f}$, and average pairing gaps are listed in Table~\ref{tab:edf_comparison}. Unlike Fig.~\ref{fig:mass1}, Fig.~\ref{fig:mass2} compares the NLDs obtained with PC-PK1 and its adjusted versions with those of DD-ME2 over the excitation-energy range 0--40~MeV. It can be seen that, as the effective mass decreases, the NLDs gradually approach those of DD-ME2. The pronounced suppression exhibited by PC-PK1 at low excitation energies, as well as its enhancement at high excitation energies, is progressively weakened in PC-PK1a, PC-PK1b, and PC-PK1c. At 40~MeV, the NLD obtained with PC-PK1c is only about 1.2 times that of DD-ME2. In combination with Table~\ref{tab:edf_comparison}, this trend is consistent with the systematic changes in $g_{\varepsilon_f}$ and the average pairing gaps. It should be noted, however, that this overall trend is not fully reflected in the single-particle level densities listed in Table~\ref{tab:edf_comparison} for $\sigma=0.5$ MeV. This is because the present parameter readjustment modifies the spin-orbit potential through the self-consistent RHB iteration, thereby affecting the arrangement of the single-particle levels around the Fermi surface. As a result, $g_{\varepsilon_f}$ defined within such a narrow energy interval may exhibit a behavior different from that seen on larger energy scales. For PC-PK1a, the delicate balance between $g_{\varepsilon_f}$ and the pairing gap leads to the closest agreement with DD-ME2 in the low-energy region. Meanwhile, because its rotational contribution is stronger than that of PC-PK1, it gives a larger NLD in the range of about 20--30~MeV. A similar behavior is also seen in the comparison between PC-PK1b and PC-PK1c. 

Overall, the NLD behavior is strongly regulated by pairing effects at low excitation energies, while in the range of about 5--25~MeV it is influenced by vibrational and rotational contributions. At higher excitation energies, however, p-h excitations eventually become dominant, so that the overall ordering of the NLDs more directly reflects the differences in the single-particle level density around the Fermi surface.

\subsection{Characteristic features of the calculated NLDs}

In addition to the total NLD itself, its parity and spin distributions also provide important information for characterizing the properties of nuclear excited states, and have a direct impact on the prediction of reaction cross sections.  Since the main qualitative features of the calculated NLDs are similar for the three adopted EDFs, DD-PC1 is used in this section to illustrate the general trends. 

\begin{figure}
\centering
\includegraphics[width=8cm]{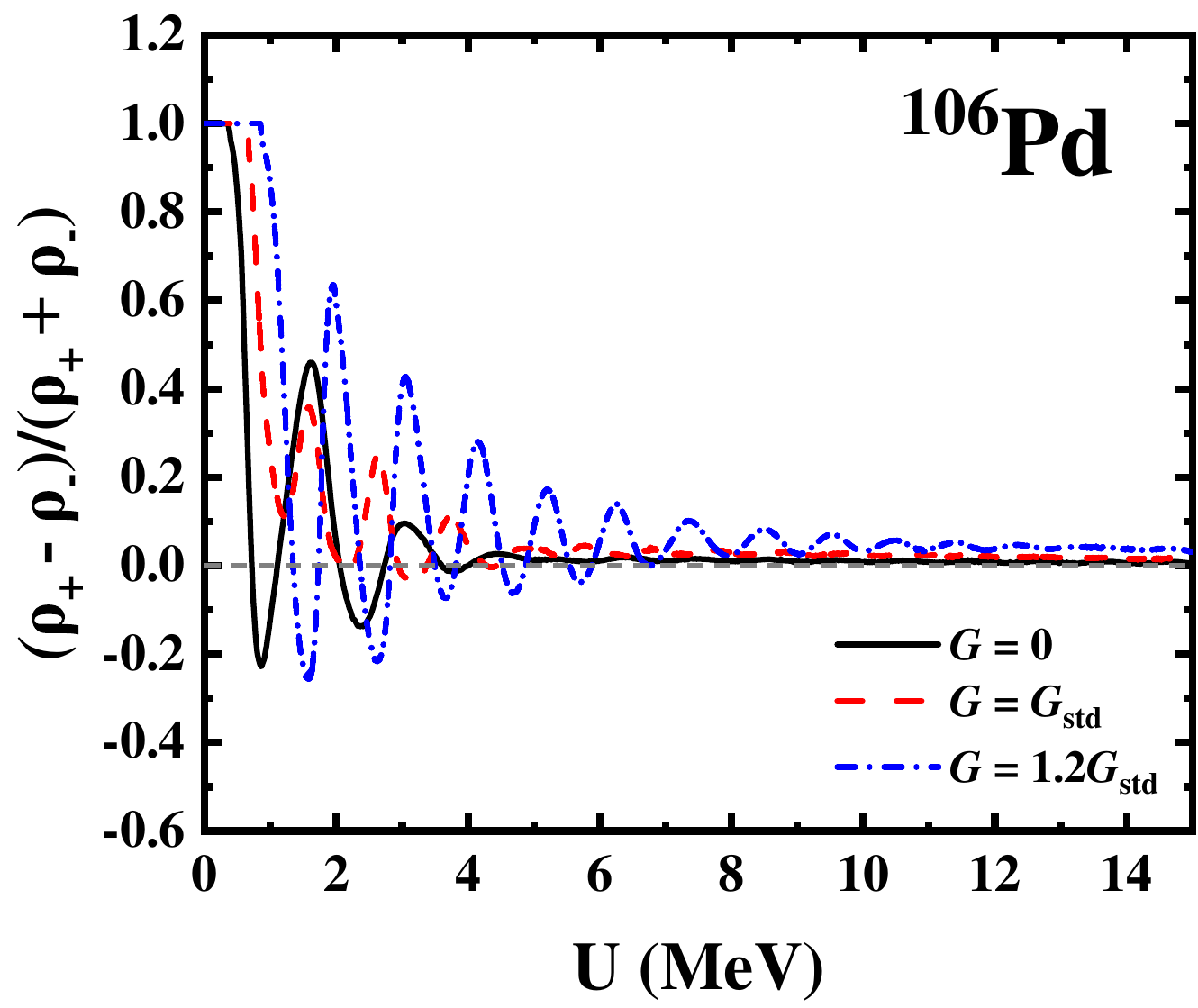}
\caption{\label{fig:parity}(Color online) 
 Asymmetry between the positive-parity ($\rho_+$) and negative-parity ($\rho_-$) NLDs for $^{\mathrm{106}}\mathrm{Pd}$, calculated with DD-PC1 using pairing strengths of 0 (black solid line), 1.0 (red dashed line), and 1.2 (blue dash-dotted line) times the adopted pairing strength.}
\end{figure}

\begin{figure}
\centering
\includegraphics[width=8cm]{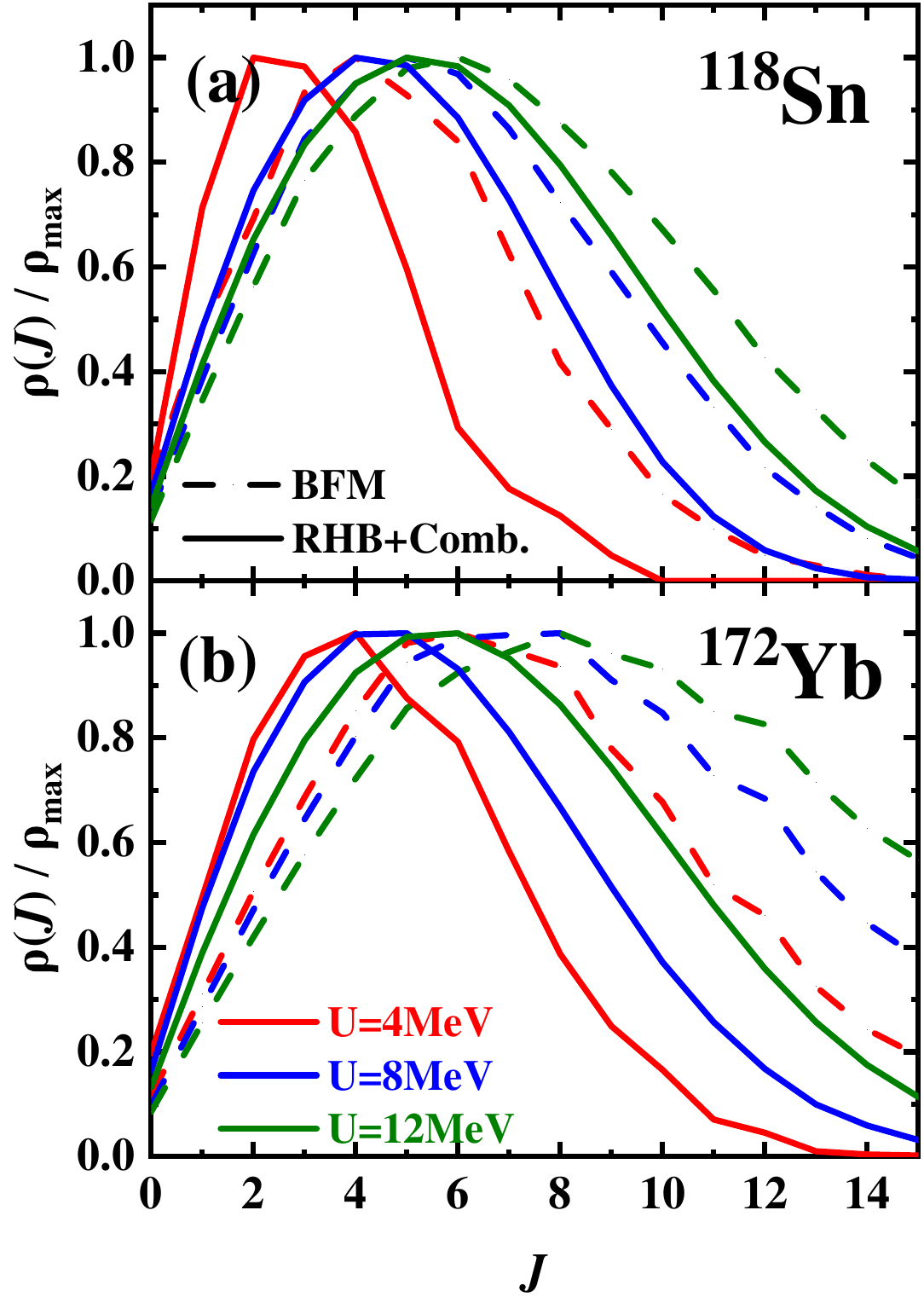}
      \caption{\label{fig:spin}(Color online) Normalized spin distributions of NLDs for the spherical $^{\mathrm{118}}\mathrm{Sn}$ (a) and the deformed $^{\mathrm{172}}\mathrm{Yb}$ (b) at three excitation energies between 4 and 12 MeV, calculated with the RHB plus combinatorial method based on DD-PC1 (solid lines), together with a comparison to the BFM results from the TALYS code~\cite{KONING2012modern} (dashed lines). }
\end{figure}

\begin{figure*}
\centering
\includegraphics[width=18cm]{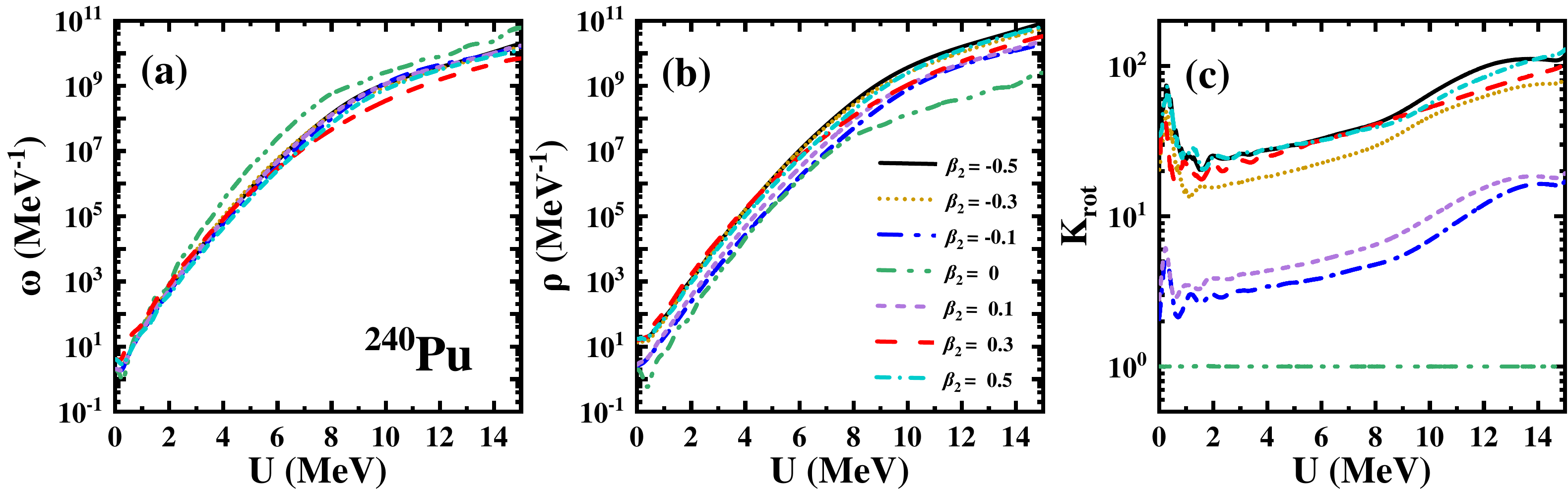}
\caption{\label{fig:def}State densities before including rotational effects (a), NLDs (b), and rotational enhancement factors ${\mathrm{K}}_\mathrm{rot}$ (c) of $^{\mathrm{240}}\mathrm{Pu}$ for different deformations with ${\beta}_2$ ranging from $-0.5$ to $0.5$, obtained with DD-PC1.}
\end{figure*}

\begin{table*}
\caption{\label{tab:pu240_beta2}%
Neutron and proton single-particle level densities near the Fermi surface $g_{\varepsilon_f}$ (in $\mathrm{MeV}^{-1}$) at different smoothing scales and the corresponding average pairing gaps $\langle uv\Delta\rangle$ (in $\mathrm{MeV}$), obtained from constrained RHB calculations with DD-PC1 for $^{240}\mathrm{Pu}$ at different deformations $\beta_{2}$.
}
\begin{ruledtabular}
\begin{tabular}{ccccccccc}
\multirow{2}{*}{$\beta_2$}
& \multicolumn{2}{c}{$g_{\varepsilon_f}^{\sigma=0.5}$}
& \multicolumn{2}{c}{$g_{\varepsilon_f}^{\sigma=1.0}$}
& \multicolumn{2}{c}{$g_{\varepsilon_f}^{\sigma=2.0}$}
& \multicolumn{2}{c}{$\langle uv\Delta\rangle$} \\
\cline{2-3}
\cline{4-5}
\cline{6-7}
\cline{8-9}
& $n$ & $p$
& $n$ & $p$
& $n$ & $p$
& $n$ & $p$ \\
\colrule
-0.5 & 5.446 & 5.817 & 6.505 & 4.618 & 6.194 & 4.254 & 0.876 & 1.009 \\
-0.3 & 6.799 & 4.058 & 6.465 & 4.160 & 5.965 & 4.175 & 0.879 & 0.956 \\
-0.1 & 6.531 & 2.078 & 7.504 & 3.263 & 6.926 & 4.227 & 1.009 & 0.773 \\
 0.0 & 12.411 & 1.353 & 9.405 & 4.330 & 7.327 & 4.658 & 1.167 & 0.884 \\
 0.1 & 7.259 & 4.409 & 7.703 & 4.146 & 6.699 & 4.149 & 1.021 & 0.945 \\
 0.3 & 4.299 & 2.277 & 3.719 & 3.157 & 4.670 & 3.895 & 0.560 & 0.768 \\
 0.5 & 4.559 & 5.598 & 5.962 & 4.850 & 6.009 & 4.355 & 0.780 & 1.040 \\
\end{tabular}
\end{ruledtabular}
\end{table*}

Figure~\ref{fig:parity} takes $^{106}\mathrm{Pd}$ as an example and shows the parity distribution calculated with DD-PC1 under different pairing strengths. The vertical axis characterizes the asymmetry between the positive-parity and negative-parity NLDs, and values closer to zero indicate that the parity distribution is closer to equilibrium. It can be seen that the parity asymmetry is pronounced at low excitation energies and gradually decreases as the excitation energy increases, indicating that more states with different parities are excited and the positive-parity and negative-parity distributions tend toward equilibrium. At the same time, pairing effects have a significant influence on the establishment of parity equilibrium by suppressing low-energy excitations. For the case of 1.2 times the standard pairing strength, the decay of the parity asymmetry is the slowest, indicating that parity equilibrium is reached the latest; in contrast, when pairing correlations are not included, the positive-parity and negative-parity distributions approach equilibrium at lower excitation energies, while the result for the standard pairing strength lies in between. This also demonstrates the importance of pairing effects in the description of NLDs.

Figure~\ref{fig:spin} shows the normalized spin distributions of the spherical nucleus $^{118}\mathrm{Sn}$ and the deformed nucleus $^{172}\mathrm{Yb}$ at three excitation energies, calculated with the RHB plus combinatorial method based on DD-PC1, together with a comparison to the widely used BFM results. It can be seen that, as the excitation energy increases, the spin distributions of both nuclei gradually broaden and extend toward higher angular momenta. At the same time, clear differences can be observed between the spherical and deformed cases. For $^{118}\mathrm{Sn}$, the spin distribution is relatively narrow, with its peak mainly concentrated in the low-spin region. In contrast, $^{172}\mathrm{Yb}$ shows a broader spin distribution, mainly because of its larger p-h space and the presence of rotational degrees of freedom. In addition, compared with the BFM, the RHB plus combinatorial method gives a narrower distribution, and this feature is more pronounced for $^{172}\mathrm{Yb}$.

\begin{figure*}
\centering
\includegraphics[width=18cm]{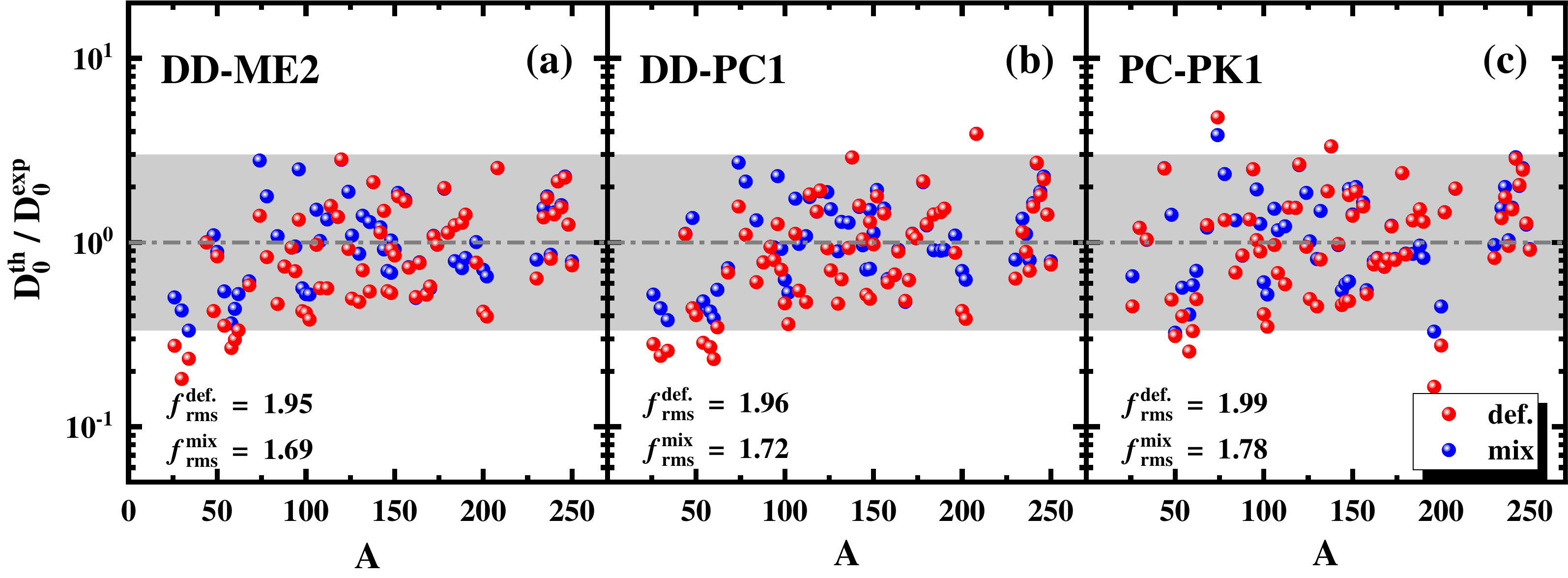}
\caption{\label{fig:d0}(Color online) Ratios of the calculated $s$-wave neutron resonance spacings ($D_{0}$) to the experimental values for DD-ME2 (a), DD-PC1 (b), and PC-PK1 (c). The red dots represent the results obtained by considering only the nuclear ground-state deformation, while the blue dots represent the results obtained by mixing the ground-state deformation and spherical cases through a phenomenological damping function.}
\end{figure*}

The influence of deformation on NLDs is further discussed here. In the previous subsection, this effect is illustrated by the differences in the NLDs of $^{74}\mathrm{Ge}$ associated with the different ground-state deformations predicted by different EDFs. A more detailed analysis is presented for $^{240}\mathrm{Pu}$ based on DD-PC1. For $\beta_2$ ranging from $-0.5$ to $0.5$, the state densities before including rotational effects, the NLDs, and the corresponding rotational enhancement are shown in Fig.~\ref{fig:def}. As can be seen from Fig.~\ref{fig:def}(a), the state densities are similar for most deformations, whereas the cases of $\beta_2=0.0$ and $\beta_2=0.3$ show markedly different behavior. In the spherical case, the state density is smallest at low excitation energies but rises rapidly and eventually becomes the largest, while the opposite behavior is found for $\beta_2=0.3$. At around 15~MeV, the difference between the two already approaches one order of magnitude. Table~\ref{tab:pu240_beta2} lists $g_{\varepsilon_f}$ near the Fermi surface at different smoothing scales, together with the corresponding average pairing gaps, for different deformations. On the local scales of $\sigma=0.5$ and $1.0$, the differences in $g_{\varepsilon_f}$ are significant and are reflected in average pairing gaps. In particular, for the spherical case, the proton $g_{\varepsilon_f}$ at $\sigma=0.5$ is remarkably small, which can be associated with the spurious shell closure around $Z=92$~\cite{Zhao2022Covariant}. At larger smoothing scales, the differences in $g_{\varepsilon_f}$ gradually decrease. For most deformations, the opposite effects of $g_{\varepsilon_f}$ and pairing gap on the state density keep the differences below 15~MeV relatively small. The spherical shape and the prolate deformation with $\beta_2=0.3$, however, correspond to particularly large and small values of $g_{\varepsilon_f}$, respectively. The former mainly results from the higher level degeneracy of the spherical case, whereas the latter reflects the fact that $\beta_2=0.3$ is already close to the ground-state deformation of $^{240}\mathrm{Pu}$ $(\beta_2=0.28)$, where the system tends to minimize its energy by reducing the single-particle level density around the Fermi surface.

After rotational effects are taken into account, the differences among the results for different deformations are further amplified, as shown in Fig.~\ref{fig:def}(b). At the same time, the rotational enhancements extracted from the corresponding state densities are displayed in Fig.~\ref{fig:def}(c). It can be seen that the rotational enhancement remains relatively weak for small deformations, whereas it increases rapidly when the system develops more pronounced prolate or oblate shapes. This behavior is associated with the larger moment of inertia and leads to a significant enhancement of the NLD. In addition, these rotational enhancements exhibit some common features. A peak appears around 0.2~MeV, which mainly originates from the fact that the local minimum formed near the ground state after smoothing with Eq.~\ref{eq:formula_smooth} is filled by rotational levels. Another maximum appears around 14--15~MeV, which is mainly related to the vibrational enhancement. This maximum decreases gradually within a certain excitation-energy range as the vibrational effect fades away. At still higher excitation energies, however, the rotational contribution increases again, since an explicit fade-out mechanism for the rotational enhancement is not yet included in the present framework.

\subsection{Global assessment through $s$-wave neutron resonance spacings}

The $s$-wave neutron resonance spacing $D_0$ near the neutron separation energy $S_n$ provides one of the most important experimental constraints on the NLD. For a target nucleus with ground-state spin $J_t$ and parity $\pi_t$, $D_0$ is related to the NLD of the compound nucleus formed by $s$-wave neutron capture as follows:
\begin{equation}
\frac{1}{D_0}
=
\sum_{J_f = |J_t \pm 1/2|}
\rho\!\left(S_n,J_f,\pi_t\right),
\label{eq:d0rho}
\end{equation}
where the accessible spins are determined by the coupling of the target spin with the neutron spin in the $s$-wave channel. On this basis, the overall performance of the RHB plus combinatorial method based on the three relativistic EDFs is assessed in the present work using 67 even-even nuclei with available experimental $D_0$ data from the RIPL-3 database~\cite{capote2009ripl}. The red dots in Fig.~\ref{fig:d0} show the ratios of the calculated $D_0$ values to the experimental data. These results are obtained with the combinatorial method, using the ground-state information predicted by the three relativistic EDFs. To quantify the overall deviation between theory and experiment, the root-mean-square deviation factor $f_{\mathrm{rms}}$ is adopted as a global measure~\cite{hilaire2001combinatorial}:
\begin{equation}
f_{\textrm{rms}}=\mathrm{exp}{\left[\frac{1}{N}\sum^{N}_{i=1}{\ln^2\frac{D^{th}_{0}}{D^{exp}_{0}}}\right]}^{1/2},
\label{eq:rms}
\end{equation}
where $N$ is the number of nuclei considered. The resulting $f_{\mathrm{rms}}$ values for the three EDFs are also listed in Fig.~\ref{fig:d0}. It can be seen that, for all three EDFs, the deviations from the experimental data are generally within a factor of two. The corresponding $f_{\mathrm{rms}}$ values are 1.95, 1.96, and 1.99, respectively, which are already comparable to those of the most successful global NLD models currently available~\cite{demetriou2001microscopic,goriely2008improved1,Goriely2026Improved}. For nuclei with mass number below 70, the three EDFs generally underestimate the experimental $D_0$ values, which implies an overestimation of the NLDs, and this tendency is weaker for PC-PK1. However, for heavy nuclei with mass number above 150, PC-PK1 tends to predict larger $D_0$ values than the other two EDFs. This behavior is precisely a consequence of the different single-particle level densities and pairing properties associated with different effective masses. Meanwhile, for nuclei near shell closures, all three EDFs tend to give larger $D_0$ values, corresponding to an underestimation of the NLDs. This behavior may be associated with the relatively low effective masses of relativistic EDFs and the resulting sparse single-particle spectra, which can enhance shell effects in the vicinity of shell closures, and may also be affected by the spin and parity distributions of the NLD.  

The present framework provides a good description of NLDs for well-deformed nuclei, and the treatment of spherical nuclei is also rather clear. However, for weakly deformed nuclei, the treatment of rotational effects may lead to an overestimation of the NLD. In addition, the present framework cannot explicitly describe phenomena such as the shape transition with increasing temperature and the fade-out of rotational enhancement. In previous combinatorial studies, this issue has often been improved by introducing a phenomenological damping function to mix the NLDs obtained for the ground-state deformation and for the spherical configuration of the same nucleus~\cite{hilaire2006global,goriely2008improved1,Goriely2026Improved}. Its form is written as
\begin{equation}
\rho(U,J,\pi)
=
\mathcal{F}\rho_{\mathrm{def}}(U,J,\pi)
+
\left[1-\mathcal{F}\right]\rho_{\mathrm{sph}}(U,J,\pi),
\label{eq:shape_mixing}
\end{equation}
where $\rho_{\mathrm{def}}$ and $\rho_{\mathrm{sph}}$ denote the NLDs calculated for the ground-state deformation and the spherical configuration, respectively, and $\mathcal{F}$ is the phenomenological damping function. To avoid introducing additional phenomenological degrees of freedom and to more clearly examine the direct influence of different EDFs, the mixing treatment is not adopted in the main results presented above. For reference, the $D_0$ values obtained by applying this mixing prescription are shown by the blue dots in Fig.~\ref{fig:d0}. The damping function $\mathcal{F}$ is taken to be energy dependent,
\begin{equation}
\mathcal{F}(U)
=
\frac{1}{1+e^{(U-E_{\mathrm{diff}}-E^{*})/dU}},
\label{eq:fdamp}
\end{equation}
where $E_{\mathrm{diff}}$ denotes the total energy difference between the spherical configuration and the ground-state configuration, while $E^{*}$ and $dU$ are parameters controlling the damping behavior, both given in MeV. With this prescription, the fitted $f_{\mathrm{rms}}$ values become 1.69, 1.72, and 1.78 for DD-ME2, DD-PC1, and PC-PK1, respectively. The corresponding parameter sets $(E^{*},dU)$ are chosen as $(7.39,2.88)$, $(7.54,3.62)$, and $(8.35,3.46)$. As shown by the blue dots in Fig.~\ref{fig:d0}, the improvement mainly appears for lighter nuclei and weakly deformed nuclei, such as $^{200}\mathrm{Hg}$. At the same time, the remaining deviations suggest that a more microscopic treatment of temperature-dependent shape evolution and the fade-out of collective enhancement would be important for further improving the description, especially for weakly deformed nuclei and nuclei near shell closures.

\section{\label{sec:4}SUMMARY AND PROSPECTS}

In this work, the NLDs of 67 even-even nuclei have been systematically investigated within the relativistic Hartree-Bogoliubov plus combinatorial framework using the relativistic energy density functionals DD-ME2, DD-PC1, and PC-PK1. The performance of the method has been examined through comparisons with known low-lying levels, Oslo data, and $s$-wave neutron resonance spacings $D_0$.

The results show that the three relativistic EDFs provide a good description of the available experimental data. In the energy region covered by the Oslo data, the calculations reproduce well the overall increase of the experimental NLDs with excitation energy and, to some extent, retain the nonstatistical features in the low-energy region. The results obtained with different functionals are generally close to each other, but systematic differences can still be observed. These differences mainly originate from the single-particle level structure around the Fermi surface and the pairing properties predicted by different functionals, which are closely related to the effective masses at nuclear matter saturation. The additional adjustment of the PC-PK1 parameters also shows that, as the effective masses approach those of DD-ME2, the calculated NLDs move correspondingly closer to the DD-ME2 results.

The parity distribution, spin distribution, and deformation dependence of the NLDs have also been discussed. The results show that the parity asymmetry mainly appears in the low-excitation-energy region and gradually weakens with increasing excitation energy. Pairing correlations delay the establishment of parity equilibrium by suppressing low-energy excitations. The spin distributions become broader with increasing excitation energy, but are evidently narrower than those given by the BFM. Deformation affects the final NLD not only by changing the single-particle level structure around the Fermi surface and the pairing properties, but also through its significant influence on rotational enhancement.

Finally, a global assessment has been performed using 67 even-even nuclei with available experimental $D_0$ data from the RIPL-3 database. The root-mean-square deviation factors $f_{\mathrm{rms}}$ of the calculated $D_0$ values with respect to the experimental data are 1.95, 1.96, and 1.99 for DD-ME2, DD-PC1, and PC-PK1, respectively, indicating that the accuracy of the RHB plus combinatorial method in describing $s$-wave neutron resonance spacings is comparable to that of successful existing global NLD models. As a reference, after introducing a phenomenological mixing of ground-state deformation and spherical configurations, the corresponding $f_{\mathrm{rms}}$ values are reduced to 1.69, 1.72, and 1.78, with the improvement mainly coming from light nuclei and weakly deformed nuclei.

Overall, the RHB plus combinatorial method provides reliable spin- and parity-dependent NLDs within the relativistic EDF framework and gives a good description of the available experimental information. Further improvements may require a more self-consistent treatment of the temperature dependence of shape evolution and pairing correlations, the fade-out of collective enhancement, and additional deformation degrees of freedom.

\begin{acknowledgments}
This work is supported by the National Natural Science Foundation of China (No.12475119) and Key Laboratory of Nuclear Data Foundation (JCKY2025201C154).
\end{acknowledgments}

\nocite{*}

\bibliography{apssamp}

\end{document}